\documentclass[twocolumn,prl,showpacs]{revtex4}
\usepackage{graphicx}
\usepackage{bm}
\newcommand\Gd{Gd$^{3+}$}

\newcommand\lll{$H \! \parallel \! [111]$}
\newcommand\llo{$H \! \parallel \! [110]$}
\newcommand\llt{$H \! \parallel \! [112]$}
\newcommand\GTO{Gd$_2$Ti$_2$O$_7$}
\newcommand\PRB[3]{Phys. Rev. B {\bf {#1}}, {#2} ({#3})}
\newcommand\ibid[3]{{\it ibid.} {\bf {#1}}, {#2} ({#3})}
\newcommand\PRL[3]{Phys. Rev. Lett. {\bf {#1}}, {#2} ({#3})}
\newcommand\JPCM[3]{J. Phys. Condens. Matter {\bf {#1}}, {#2} ({#3})}
\newcommand\Tno{$T_{N1}$}
\newcommand\Tnt{$T_{N2}$}

\begin{document}
\title{Magnetic phase diagram of the antiferromagnetic pyrochlore \GTO}
\author{O.~A.~Petrenko, M.~R.~Lees, G.~Balakrishnan and D.~$\rm M^cK$~Paul}
\affiliation{University of Warwick, Department of Physics, Coventry, CV4~7AL, UK}
\date{\today}
\begin{abstract}
\GTO\ is a highly frustrated antiferromagnet on a pyrochlore lattice, where apart from the Heisenberg exchange the spins also interact via dipole-dipole forces. We report on low-temperature specific heat measurements performed on single crystals of \GTO\ for three different directions of an applied magnetic field. The measurements reveal the strongly anisotropic behaviour of \GTO\ in a magnetic field despite the apparent absence of a significant single-ion anisotropy for \Gd. The $H-T$ phase diagrams are constructed for \lll, \llo\ and \llt. The results indicate that further theoretical work beyond a simple mean-field model is required.
\end{abstract}

\maketitle

Amongst the members of the titanium pyrochlore oxide family with the general formula R$_2$Ti$_2$O$_7$, the magnetic properties of which are all presently the subject of intense investigations, gadolinium titanium oxide (GTO) holds a unique position. In contrast to the other members of this series of compounds, where magnetic anisotropy plays a crucial role and results in the appearance of some most unusual ground state configurations, including the ``spin-ice" state~\cite{Bramwell_Science01}, GTO is the only compound where the magnetic rare earth ions have a negligible single-ion anisotropy. This is because in \GTO, the \Gd\ ions are in a state with $S=7/2$ and $L=0$. The magnetic ions are antiferromagnetically coupled. Therefore, apart from the ever present dipole-dipole interactions, GTO may be regarded as a realisation of an ideal Heisenberg antiferromagnet on a frustrated pyrochlore lattice.

The considerable theoretical interest in the 3$d$ pyrochlore model and its 2$d$ analogue (a checkerboard lattice), is mostly due to the fact that these models, both in classical and quantum descriptions, do not develop conventional magnetic order down to $T=0$~K~\cite{Pyro_General}. Their ground state is instead described as a degenerate spin-liquid~\cite{Spin_Liquid}. Whether the weaker perturbations such as a single ion anisotropy, dipolar interactions~\cite{Dipolar,Dipolar_Enjalran}, small lattice distortions~\cite{Distortions}, further-neighbor interactions~\cite{NNN} or a magnetic field~\cite{Field} break the degeneracy and force the system to adopt a particular ground state configuration is the subject of current investigations. In this context it is extremely important to identify different magnetic phases in real magnetic materials and to establish the phase diagrams.

Another unusual aspect of the thermodynamics of frustrated magnets, which may have direct consequences for GTO, has been pointed out recently by Zhitomirsky~\cite{MZH_PRB}. Because a significant enhancement of the magnetocaloric effect is expected for frustrated magnets near the saturation field, GTO may have potential as an adiabatic demagnetisation refrigerator. An enhancement of the magnetocaloric effect in model systems is attributed to the presence of a significant number of so-called ``soft modes"~\cite{MZH_PRB}. The existance of such soft (and zero) modes in a magnetic material such as \GTO, still needs to be established. To date, the only direct experimental observation of soft modes has been made with inelastic neutron scattering in another frustrated compound ZnCr$_2$O$_4$~\cite{ZnCrO}.

In this letter, we present the results of specific heat measurements on \GTO\ in an applied magnetic field. By using single crystal samples we have been able to prove that there is an unusually high degree of anisotropy built into the ground state and to construct the $H-T$ phase diagrams for three different directions of magnetic field, \lll, \llo\ and \llt.

The single crystals of \GTO\ were grown by the floating zone technique, using an infrared image furnace~\cite{Balakrishnan}. The principal axes of the samples were determined using X-ray diffraction Laue photographs; the crystals were aligned to within an accuracy of 2-4$^{\circ}$. The specific heat measurements were performed in the temperature range 0.39 to 5.0~K in a field of up to 9~T using a Quantum Design PPMS calorimeter equipped with a $^3$He option. In order to estimate the lattice contribution to the specific heat we have measured the specific heat of a single crystal of Y$_2$Ti$_2$O$_7$, a nonmagnetic compound isostructural to GTO. For all the low-temperature measurements reported in this letter ($T<5$~K) the lattice contribution was found to be insignificant. 

We have measured the specific heat both as a function of temperature in a constant magnetic field and as a function of applied field at constant temperature. Because of the relatively high values of the specific heat of GTO and its poor thermal conductivity at low temperatures, the measurements were restricted to small ($<1$ mg) plate-like samples with a typical thickness of 0.1 mm. For some orientations of applied field, the demagnetizing factors were significant and had to be taken into account~\cite{demag}. This has been done by measuring the magnetization of the same or similarly shaped samples with either a vibrating sample or a SQUID magnetometer.

\begin{figure}
\includegraphics[width=0.95\columnwidth]{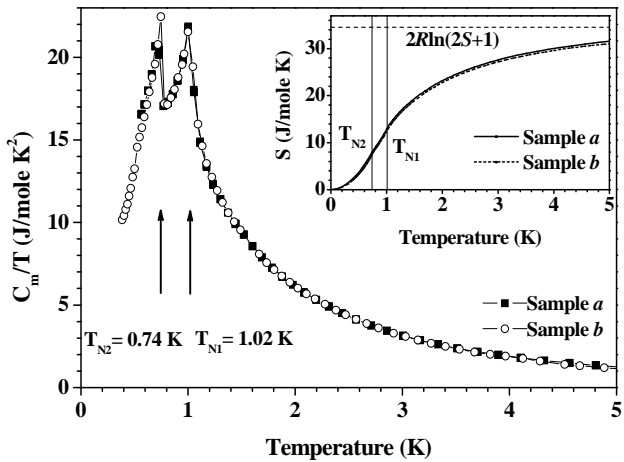}
\caption{\label{Fig1} Temperature dependence of the specific heat  divided by temperature, $C(T)/T$, measured on the single crystal of the \GTO\ in a zero applied field. The inset shows the temperature dependence of the magnetic entropy.}
\end{figure}

The temperature dependence of the heat capacity divided by temperature, $C(T)/T$, measured in zero field is shown in Fig.~\ref{Fig1}. Two sharp peaks of nearly the same amplitude observed at \Tno =1.02(2)~K and \Tnt =0.74(2)~K mark the temperature of two consecutive phase transitions. In Fig.~1 two data sets are shown in order to demonstrate the high degree of reproducibility in these measurements. These data should be compared with the results of previous measurements performed on powder samples~\cite{Raju,Ramirez}. Raju {\it et al.}~\cite{Raju} reported a broad peak centred around 2~K (absent in our data) and a sharp peak just below 1~K, which were attributed to the presence of short range correlations and the development of long range magnetic order respectively. More recently, however, Ramirez {\it et al.}~\cite{Ramirez} have reported the presence of another sharp peak at around 0.7~K. While our data are similar to those presented in Ref. \cite{Ramirez}, they differ significantly in the absolute values of $C(T)$ from the values reported in Ref. \cite{Raju}. For example, at the temperature of the upper phase transition, \Tno =1.02~K in our data, the specific heat reaches a value of 22~${\rm J mol^{-1}K^{-1}}$, while the values reported in Ref.~\cite{Raju} are approximately half this quantity. Moreover, Bonville {\it et al.} \cite{Bonville} have very recently repeated the measurements of specific heat on a powder sample of GTO and reported values which are in a good agreement with our data.

In terms of the overall magnetic entropy, Raju {\it et al.} observed a significant magnetic specific heat up to about 30~K, while Ramirez {\it et al.} have recovered 92\% of the expected entropy, $2R\ln(2S+1)=34.6 \; {\rm J mol^{-1}K^{-1}}$, between 0 and 5~K. The differences in the specific heat data, as well as in the $ac$ magnetic susceptibility behaviour, have been suggested to arise from the extreme sensitivity of the low-temperature properties to the sample preparation technique~\cite{Ramirez}. The inset in Fig.~1 shows the temperature dependence of the magnetic entropy, obtained by extrapolating the $C(T)/T$ to 0 K and numerically integrating it versus temperature. Approximately 90\% of the expected value is recovered when the temperature reaches 5 K. Additional high temperature measurements have shown that the entropy is almost fully recovered by 10 K, a temperature which corresponds to the reported Curie-Weiss temperature of GTO~\cite{Raju}. At that point the magnetic contribution to the specific heat becomes negligible when compared to the lattice contribution. The fact that a majority of entropy is recovered only at $T>$\Tno\ signifies the importance of short-range correlations in GTO.

\begin{figure}
\includegraphics[width=0.95\columnwidth]{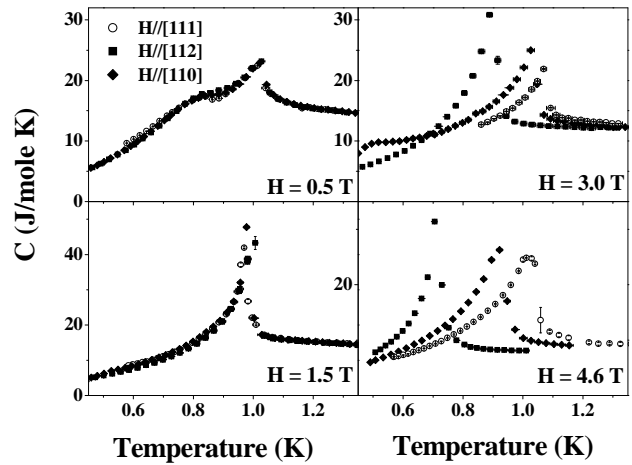}
\caption{\label{Fig2} Temperature dependence of the specific heat, $C(T)$, of \GTO\ at various values of an external magnetic field applied along the three different crystallographic directions, \lll, \llt\ and \llo.}
\end{figure}

Powder neutron diffraction measurements performed on GTO sample by Champion {\it et al.} in zero field~\cite{Champion}, have revealed a complex noncollinear structure with a magnetic propagation vector of ${\bf k}=(\frac{1}{2}\frac{1}{2}\frac{1}{2})$. According to these measurements, the GTO magnetic structure (the so-called P state) consists of kagome planes with ``$q=0$"-type order with interstitial sites carrying zero magnetic moment; GTO has been suggested to be only partially ordered even at 50~mK~\cite{Champion}. The observed recovery of the full magnetic entropy conflicts with this suggestion. On the other hand, mean-field theory~\cite{Dipolar_Enjalran} predicts that an ordered four sublattice structure (F state) competes with the P state and becomes more stable as the temperature is reduced. Therefore we suggest that at  \Tno\ GTO undergoes a transition from a paramagnetic to the P state, while at  \Tnt\ the transition is to the F state.

Our lowest experimentally accessible temperature of 0.39~K is about 50\% of \Tnt. We cannot  therefore reliably extrapolate the $C(T)$ dependence down to 0~K. From the data shown in Fig.~1, however, it is quite clear that between 0.39~K and \Tnt, $C(T)/T$ grows approximately linearly with temperature, which implies a $T^2$ dependence for the low-temperature part of $C(T)$. A similar $T^2$ dependence has been observed in another magnetically frustrated system, $\rm SrCr_{9p}Ga_{12-9p}O_{19}$ (SCGO) with a kagome lattice \cite{SCGO}.

\begin{figure}
\includegraphics[width=0.95\columnwidth]{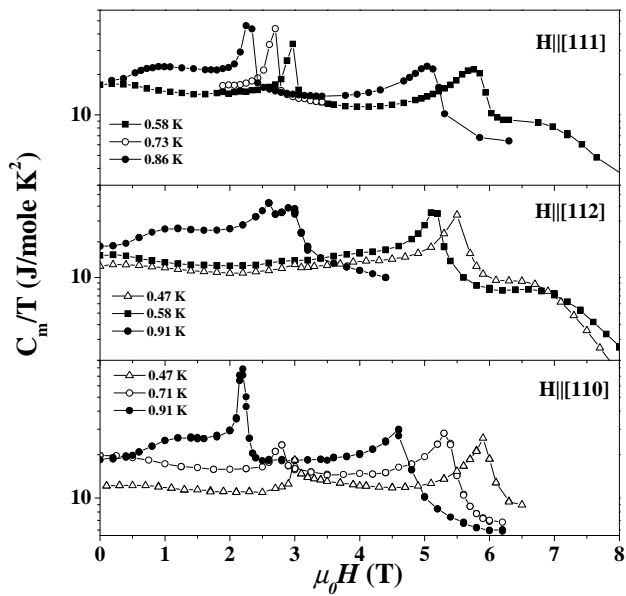}
\caption{\label{Fig3} Field dependence of the \GTO\ specific heat divided by temperature, $C(T)/T$, for three different directions of an applied magnetic field.}
\end{figure}

The temperature and field dependencies of the specific heat, measured on the single crystals of GTO for the three different directions of applied magnetic field, \lll, \llt\ and \llo\ are shown in Figures 2 and 3. The application of a small external field causes the convergence of the transition temperatures, \Tno\ and \Tnt, regardless of the field orientation. In a field of about 1.5~T, the two transitions merge~\cite{coincide}, as seen on the left-hand panels of Fig.~2. In a small field and low-$T$, $C(T)$ still follows a $T^2$ dependence; above 1.5~T, however, there is no $T^2$ dependence in  $C(T)$. This is in contrast with SCGO, whose specific heat shows almost no field dependence \cite{SCGO}. For fields above 1.5~T, significant differences in the positions of the peaks in the $C(T)$ curves for GTO are clearly visible (see right-hand panels of Fig.~2).

The field dependence of the specific heat shown in Fig.~3 has been measured in order to determine precisely the positions of the phase transitions lines at low $T$, where in high fields the specific heat is nearly temperature independent. The higher-field peaks observed in these $C(H)$ curves at around 5 to 6~T, mark a transition to a saturated (paramagnetic) phase. The low-$T$ peaks found at approximately half the saturation field correspond to an additional field-induced transition discussed below. The most intense peaks in $C(H)$ are found in the higher-$T$ low-field region. They reach values above 70~${\rm J mol^{-1}K^{-1}}$ and correspond to a transition into an intermediate temperature phase. It is interesting to compare this $C(H)$ data with the results of Monte Carlo simulations for classical spins on the pyrochlore lattice near the saturation field~\cite{MZH_PRB}, where the presence of a macroscopic number of soft modes has been predicted. This comparison is not straightforward, as in the simulations only the exchange interactions have been included, which precludes the system from entering a state with long range magnetic order~\cite{MZH_PRB}.
Fig.~3 shows that for $H>H_{sat}$ the measured field dependence of the specific heat does not follow a simple exponential decay expected for a nonfrustrated antiferromagnet. Instead it remains almost field-independent for $H<7$~T and then falls rapidly at higher fields. This observation may suggest a large influence of the soft modes on the specific heat of GTO around the saturation field.

No signs of history dependence have been detected throughout the course of our measurements, {\it e.g.} the values of the measured specific heat were identical for increasing/decreasing field and temperature. This suggests a second order nature for all the phase transitions in GTO in contrast with the observations of Bonville {\it et al.}~\cite{Bonville}, which imply the upper phase transition at \Tno\ has a weak first order character .

\begin{figure}
\includegraphics[width=0.95\columnwidth]{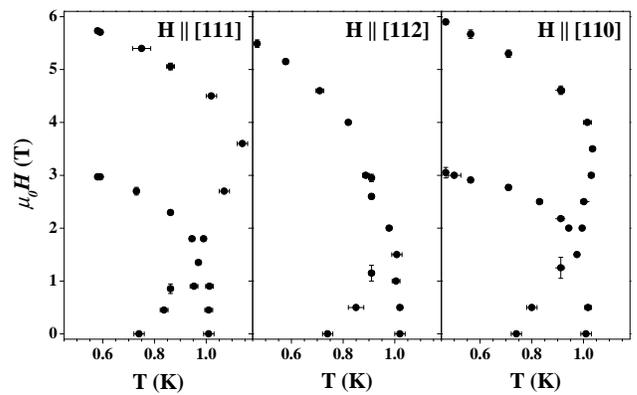}
\caption{\label{Fig4} Magnetic phase diagrams of the \GTO\ for three different orientations of an applied magnetic field.}
\end{figure}

The magnetic $H-T$ phase diagrams for GTO obtained by plotting the temperature of various phase transitions observed in both the $C \;vs.\; H$ and $C \;vs.\; T$ measurements are shown in Fig.~4. By analysing these phase diagrams, the highly anisotropic nature of the magnetic behaviour of GTO becomes more apparent. To start with, when \lll, the magnetic field favors the onset of magnetic order in GTO. For example, in a field of  about 4~T the transition temperature is 1.14(2)~K compared to 1.02(2)~K in zero field, while the transition to a saturated phase measured at 0.57~K occurs at about 5.7~T. When \llt, an external field inhibits the onset of long range magnetic order: at 4~T the transition is reduced to 0.82~K and the saturation field at 0.57~K does not exceed 5.2~T.  Application of the field along the $[110]$ direction has an intermediate effect.

Even more remarkably, an additional phase transition induced by the external field at about half of the saturation field is clearly present at low temperatures for \lll, but is absent for \llt. For \llo, this phase transition is present, but it becomes less and less pronounced as the temperature decreases (see bottom panel of Fig.~3). The origin of this phase transition is not clear at present, although the fact that it happens at a half of $H_{sat}$ strongly suggests that it is likely to be related to a collinear spin state, where three spins on each tetrahedra are aligned with the field and the fourth is pointing in the opposite direction. A stabilisation of the collinear phase by quantum and thermal fluctuations has been predicted to take place at $H=H_{sat}/2$ for Heisenberg antiferromagnets on pyrochlore and frustrated square lattices, as well as at $H=H_{sat}/3$ for kagome and garnet lattices \cite{Field}. Whether or not GTO, whose magnetic interactions are more complex than a simple Heisenberg model, does indeed adopt a collinear order in a magnetic field remains to be established with more direct probes of the magnetic structure, such as neutron diffraction measurements. 

When compared to the phase diagram obtained from the $C(H,T)$ measurements on a powder sample \cite{Ramirez}, Fig.~4 demonstrates the advantages of using single crystals. While it has been claimed that there are as many as four different ordered phases in GTO \cite{Ramirez}, our measurements show that for any given direction of magnetic field there are at most three ordered phases. The phase diagram reported by Ramirez {\it et al.} \cite{Ramirez} seems to consist of a superposition of the single crystal phase diagrams.

When compared to the theoretical predictions based on a simple mean-field model \cite{Ramirez}, our phase diagrams reveal significant discrepancies. For example, for the field along $[111]$, theory predicts only one transition, while our measurements clearly show two. These observations strongly suggest the need for further development of the theoretical model, which most likely should include some degree of magnetic anisotropy.

Recently Hassan {\it et al.}~\cite{ESR} have reported the presence of a surprisingly large anisotropy in GTO below a temperature of 80~K, which they observed by measuring the ESR spectra. The anisotropy reveals itself as a well pronounced shift in the position of one of the resonance lines from its maximum value for \lll\ to its minimum for $H \! \perp \! [111]$. The shift amounts to 3.6~T when measured at a frequency of 54~GHz. No change in the position of the resonance lines was observed when the field was rotated in a plane perpendicular to the $[111]$ direction~\cite{ESR}. The same authors have also reported {\it dc}-torque measurements \cite{dc-torque}, which have shown a gradual development of macroscopic anisotropy below 80~K.

Motivated by these results, we have measured the temperature dependence of the magnetic susceptibility of a GTO crystal for \lll\ and $H \! \perp \! [111]$ using a SQUID magnetometer. No significant anisotropy in the susceptibility has been detected \cite{SQUID}. At present, neither the origin of the anisotropy observed in the ESR measurements \cite{ESR}, nor its presence at a relatively high temperature have a satisfactory explanation. Any explanation should obviously take into consideration more than a simple Heisenberg exchange interaction between the nearest neighbor spins.


\begin{thebibliography}{99}
\bibitem{demag} The estimated demagnetization factor, $n$, varied from 0.55 to 0.65 for the samples used when a field was applied perpendicular to the sample planes. $n$ was below 0.05 for the samples used for the in-plane measurements.
\bibitem{coincide} The technique used for the $C(T)$ measurements does not allow one to distinguish peaks in the temperature dependence if they are separated by less than 0.5\% of the temperature, which limits the accuracy of the transition point measurements at $H=1.5$~T to 5~mK.
\end{thebibliography}
\end{document}